\documentclass[conference]{IEEEtran}
\IEEEoverridecommandlockouts
\usepackage{cite}
\usepackage{amsmath,amssymb,amsfonts}
\usepackage{amsthm}
\usepackage{algorithmic}
\usepackage{graphicx}
\usepackage{textcomp}
\usepackage{xcolor}
\usepackage{changes}
\usepackage{ulem}
 \usepackage{bbm}
 \usepackage{lipsum,adjustbox}

\usepackage{hyperref}
\hypersetup{
 colorlinks,
 linkcolor={blue!100!black},
 citecolor={blue!100!black},
 urlcolor={blue!80!black}
}
\usepackage[numbers,sort]{natbib}
\newtheorem{remark}{Remark}

\newcommand{\RNum}[1]{\uppercase\expandafter{\romannumeral #1\relax}}
\usepackage{soul} %For strikesout, /st

\newcommand{\etal}{\textit{et al. }}

\newcommand{\bbF}{\mathbb{F}}

\newcommand{\bfa}{\mathbf{a}}
\newcommand{\bfb}{\mathbf{b}}
\newcommand{\bfc}{\mathbf{c}}

\newcommand{\bfe}{\mathbf{e}}

\newcommand{\bfh}{\mathbf{h}}

\newcommand{\bfk}{\mathbf{k}}
\newcommand{\bfl}{\mathbf{l}}

\newcommand{\bfp}{\mathbf{p}}

\newcommand{\bfr}{\mathbf{r}}
\newcommand{\bfs}{\mathbf{s}}

\newcommand{\bfu}{\mathbf{u}}
\newcommand{\bfv}{\mathbf{v}}
\newcommand{\bfw}{\mathbf{w}}
\newcommand{\bfx}{\mathbf{x}}
\newcommand{\bfy}{\mathbf{y}}

\newcommand{\bfB}{\mathbf{B}}

\newcommand{\bfL}{\mathbf{L}}

\newcommand{\bfR}{\mathbf{R}}

\newcommand{\bfV}{\mathbf{V}}

\newcommand{\bfX}{\mathbf{X}}
\newcommand{\bfY}{\mathbf{Y}}

\begin{document}

\title{Low Latency Cross-Shard Transactions \\in Coded Blockchain}
%
%
%
% \author{\IEEEauthorblockN{Canran Wang}
% \IEEEauthorblockA{\textit{McKelvey School of Engineering} \\
% \textit{Washington University in St. Louis}\\
% St. Louis, Missouri, USA \\
% canran@wustl.edu}
% \and
% \IEEEauthorblockN{Netanel Raviv}
% \IEEEauthorblockA{\textit{McKelvey School of Engineering} \\
% \textit{Washington University in St. Louis}\\
% St. Louis, Missouri, USA \\
% netanel.raviv@wustl.edu}}

\author{\textbf{Canran Wang} and  \IEEEauthorblockN{\textbf{Netanel Raviv}}
	\IEEEauthorblockA{
		Department of Computer Science and Engineering, Washington University in St. Louis, St. Louis 63130, MO, USA\\
		}}

\maketitle
\thispagestyle{plain}
\pagestyle{plain}

\begin{abstract}
Although blockchain, the supporting technology of Bitcoin and various cryptocurrencies, has offered a potentially effective framework for numerous applications, it still suffers from the adverse affects of the impossibility triangle. Performance, security, and decentralization of blockchains normally do not scale simultaneously with the number of participants in the~network. 

The recent introduction of error correcting codes in sharded blockchain by Li~\etal  partially settles this trilemma, boosting throughput without compromising security and decentralization. In this paper, we improve the coded sharding scheme in three ways. 
First, we propose a novel \textit{2-Dimensional Sharding} strategy, which inherently supports cross-shard transactions, alleviating the need for complicated inter-shard communication protocols. 
Second, we employ distributed storage techniques in the propagation of blocks, improving latency under restricted bandwidth.
Finally, we incorporate polynomial cryptographic primitives of low degree, which brings coded blockchain techniques into the realm of feasible real-world parameters.
\end{abstract}

\section{Introduction}
Blockchain is an append-only distributed database system with a structure of linked blocks. Data, or transactions, resides in each block on the chain, and the chain is replicated in every participating node. Periodically, a new block is proposed and propagates across the entire network. Each node verifies the transactions in the new block before linking it to the local chain, and a consensus protocol guarantees that the locally stored chains agree with each other. Further, blocks are linked with hash pointers, which ensures the information which resides in early blocks is immutable.

Blockchain's decentralization and security properties alleviate the need for a centralized trusted party, making it a promising platform for various applications, such as logistics~\cite{Logistics}, healthcare management~\cite{Health} and Internet of Things~\cite{IoT}. However, its performance is not ideal for practical use. Formally, the performance of a blockchain system is defined as a combination of \textit{throughput} and \textit{latency}. Throughput is measured by the rate in which the system processes and stores data, and latency is characterized by the required time for data to be confirmed. Bitcoin~\cite{Bitcoin} has a throughput of 7 transactions per second, and transactions are confirmed after~$10$ minutes~\cite{OnPoW}, which is orders of magnitude away from other monetary systems.

The performance of Bitcoin, as an example, is inherently limited by its design. As a distributed system, the security of Bitcoin is guaranteed by the fact that the time interval between blocks is sufficiently greater than the block transmission time~\cite{OnScaling}. In other words, in order to maintain high security level, only when the majority of nodes have received the previous block shall the new block be proposed.

As a consequence, na\"{i}vely increasing block size does boost throughput, but it would take longer for the majority of nodes to receive it, resulting in a higher latency. Likewise, shortening the time interval between blocks improves both latency and throughput but degrades security. Hence, a comprehensive redesign is required, rather than reparameterization, to improve Bitcoin's performance. Many solutions have been suggested in the literature. Bitcoin-NG~\cite{BitcoinNG} decouples the leader election from the generation of blocks, allowing a randomly selected leader to continuously generate and propagate blocks for a given time interval. ByzCoin~\cite{ByzCoin} and Hybrid Consensus~\cite{HybridConsensus} replace the individual leader in Bitcoin-NG with a committee of moderate size, and the nodes within it cooperatively verify transactions using \textit{Practical Byzantine Fault Tolerance} protocol~\cite{PBFT}. Ouroboros~\cite{Ouroboros} and Algorand~\cite{Algorand} develop leader selection mechanisms based on verifiable random functions, which improve the throughput even further and relax the requirement for computation power of nodes. Although these attempts bring improvements of throughput to some extent, they change the security model adopted in Bitcoin, and degrade the security guarantees and decentralization. 

Clearly, in order for the performance of a distributed system to scale with its size, nodes should be responsible for disjoint, rather than identical, sets of tasks. That is, different nodes should verify different transactions in parallel, leading to the novel \textit{sharding} approach. 

Sharding is a technique to horizontally partition a large database into small and easily manageable chunks called \textit{shards}~\cite{SokSharding}. In the context of blockchain, the network is sliced into multiple communities of a similar size, each individually processes a disjoint portion of data, e.g., a shard that contains transactions associated with members of a community.

Sharding addresses the performance problem. The constant and relatively small community size makes the use of traditional consensus protocols feasible, as they are designed for distributed systems with limited number of participating nodes. Hence,  high throughput and low latency are guaranteed within each community. Also, the throughput scales with the number of nodes, as additional nodes form extra communities process additional transactions.

However, these performance gains come with costs. An adversary should only corrupt a certain portion of nodes in a community, instead of in the entire network, to alter verification results and harm the overall security. In response, random community rotation is required to make community corruption a moving target for adversaries. Elastico~\cite{Elastico} enforces a complete community reassignment in the beginning of every epoch using a random number generated in the previous round. RapidChain~\cite{RapidChain} and OmniLedger~\cite{OmniLedger} swap out a portion of nodes in the beginning of an epoch.

Sharding creates a distinction between two different types of transactions; a transaction is called intra-shard if the sender and the receiver belong to the same community, and called cross-shard otherwise. In fact, most transactions in sharded-blockchains are cross-shard, and validating them requires additional communication between communities, which is an added complexity, and in return degrades the system performance. OmniLedger~\cite{OmniLedger} employs a ``lock/unlock" protocol to ensure atomicity of cross-shard transactions. RapidChain~\cite{RapidChain} builds an inter-committee routing protocol to allow inter-shard verification of cross-shard transactions.

Li \etal ~\cite{Polyshard} proposed \textit{Polyshard} to address the security issue in sharded blockchains. Polyshard formulates the verification of all transactions as a computation task to be solved distributively across all nodes. Using the novel Lagrange Coded Computing scheme (LCC)~\cite{LCC}, nodes individually compute a polynomial verification function over a \textit{coded chain} and a \textit{coded block}. Then, the verification results can be obtained from any sufficiently large set of nodes by decoding the Lagrange code.

LCC guarantees correct computation, as long as the number of adversaries and straggling nodes is below a certain threshold. In other words, the validity of every transaction can be determined even if a certain number of nodes either return wrong results or remain unresponsive. The incorporation of LCC in Polyshard enables the throughput to scale, i.e., additional transactions can be verified by additional nodes, while the computation load assigned to each node remains identical.

However, the system latency is not optimal. Polyshard requires each node to receive a block containing all transactions, whose size scales linearly with the network size, to individually perform coding. Since the transmission time in a blockchain system linearly increases with data size~\cite{OnScaling}, this requirement incurs a significant delay. Further, Polyshard mainly focuses on intra-shard transactions, and does not discuss cross-shard transactions in detail. 

Finally, Polyshard focuses on balance-check, i.e., verifying if the sender has enough balance to spend, and do not explicitly incorporate cryptographic operations, e.g., verification of digital signature and computation of hash value. The latter may lead to a high-degree verification function, which by the properties of Lagrange codes decreases the number of supported shards significantly.

To address the degree problem, Polyshard suggests to represent the verification function using an arithmetic circuit. Each layer of the arithmetic circuit consists of low-degree polynomials whose outputs serve as inputs to the polynomials in the next layer. Nodes iteratively apply coded computation to polynomials in each layer and obtain intermediate results, until the verification result computed from the highest layer is obtained. This scheme requires additional rounds of communication between nodes, and leads to longer latency.

\subsection{Our Contributions}
In this paper, we inherit the use of coded computation from Polyshard to obtain higher throughout, and comparable levels of decentralization and security guarantees with respect to uncoded (i.e., ordinary) blockchain. On top of these gains, we make the following contributions.

\begin{enumerate}

\item We propose \textit{2-Dimensional Sharding}, a new technique which partitions the transactions based on their senders and receivers, respectively. This design provides inherent support for cross-shard transactions, alleviating the need for complicated communication mechanisms. This new techniques also enables an improved \textit{collateral invalidation} rate, i.e., the maximum number of transactions that must be abandoned due to invalidation of a single transaction.
\item  Inspired by Product-Matrix MBR code~\cite{ProductMatrix}, we formulate the propagation of blocks as a node repairing process in a distributed storage system. Our scheme greatly improves latency under restricted bandwidth. 
\item We adopt the \text{unspent transaction output} (UTXO) model, and formulate the verification process as computing a polynomial function whose degree increases \textit{logarithmically} with the number of transactions on the chain. In detail, our scheme addresses the degree problem by replacing current cryptographic primitives (i.e, ECDSA, SHA256 and RIPEMD-160) by multivariate cryptography in the generation and verification of a transaction. By contrast, Polyshard suggests to represent cryptographic operations by Boolean functions, which could lead to polynomials of degree as high as the length of their inputs. 

\end{enumerate}
%Explain how the three contribution above bring coded blockchains into feasibility.

Based on these three contributions, we bring coded blockchains closer to feasibility. That is, our scheme achieves low transaction confirmation time along with scalable system throughput, and removes the boundary of shards with inherent support for cross-shard transactions. 

The rest of this paper is organized as follows. Section~\ref{section:background} introduces necessary background. Section~\ref{section:Verification} explains the coded verification scheme. Section~\ref{section:propagation} discusses the propagation of transactions in the system. Section~\ref{section:discussion} analyzes the latency, security, and the supports for cross-shard transactions.  Finally, Section \ref{section:future} suggests directions for future research.

\section{Background}\label{section:background}
\subsection{The Unspent Transaction Output (UTXO) Model}

In the {Unspent Transaction Output} (UTXO) model, value resides in transactions, instead of user accounts. A transaction has inputs and outputs. An unspent output of an old transaction serves as an input to a new transaction, incurring a value transfer between the two. The old transaction output is then invalidated, since it has been spent, and new UTXOs are created in the new transaction, serving as containers of value. 

In practice, a transaction input is a pointer which references a to-be-redeemed UTXO stored in the blockchain. A transaction output contains the amount of stored value and the intended receiver's \textit{address}, which is the hash value of her public key. Besides, the sender attaches his public key and signs the transaction with his secret key. 

To verify a transaction, the hash value computed from the public key must match the address in the referenced UTXO. Then, the signature must be valid when checked by the sender's public key. This two-step verification process guarantees the sender's possession of the public and secret keys, proves his identity as the receiver of the redeemed UTXO, and protects the integrity of the new transaction. 

Although a transaction may have multiple inputs and outputs, we adopt a simplified UTXO model in our scheme for clarity, where a transaction has exactly one input and one output, and transfers one indivisible \textit{coin}. Note that the words coin and UTXO are used interchangeably throughout this paper. 

In particular, we denote a transaction by~$\bfx=(\bfu,\bfp,\bfk,\bfs)\in \bbF_q^R$, where~$\bfu$ is used to index an UTXO in the blockchain,~$\bfp$ is the sender's public key,~$\bfa$ is the receiver's address,~$\bfs$ is sender's signature on the aforementioned three vectors, and~$\mathbb{F}_q$ is a finite field with~$q$ elements.

\subsection{Lagrange Coded Computing}
Lagrange Coded Computing~\cite{LCC} (LCC) is a recent development in the field of coded computation. The task of interest is computing a multivariate polynomial~$f(X_k)$ of degree~$\deg f$ on each of the~$K$ datasets~$\{X_1,\ldots,X_K\}$. LCC employs the Lagrange polynomial to linearly combine~$K$ datasets, generating~$N$ distinct coded dataset~$\{\widetilde{X}_1,\ldots,\widetilde{X}_N\}$ with injected computational redundancy. Each worker node evaluates~$f$ on one coded dataset, and a master node completes the computation by decoding these evaluations.

LCC achieves the optimal trade-off between resiliency, security and privacy. It tolerates up to~$S$ stragglers and~$A$ adversarial nodes, defined as working nodes that are unresponsive or return erroneous results, respectively. In addition, with proper incorporation of random keys, it also prevents the exposure of the original datasets to sets of at most T colluding workers, if the following condition holds,
$$(K+T-1)\deg~f+S+2A+1\leq N.$$

Finally, we note that Lagrange coding is a linear one, i.e., every~$\widetilde{X}_i$ is a linear combination of~$\{X_j\}_{j=1}^K$. The respective~$K\times N$ generator matrix (i.e., the matrix of coefficients) has the \textit{MDS} property, i.e., every~$K\times K$ submatrix of it is invertible.
\subsection{Multivariate Cryptography}

Multivariate cryptography is one venue in achieving postquantum security. It is based on the multivariate quadratic polynomial (MQ) problem, which is believed to be hard even for quantum computers. An MQ problem involves a system of~$m$ quadratic polynomials~$\{p^{(1)},\ldots,p^{(m)}\}$ in~$n$ variables~$\{y_1,\ldots,y_n\}$ over some finite field~$\bbF_q$, i.e.,
\begin{equation*}
\begin{aligned}
p^{(1)}(y_1,\ldots,y_n)&=\sum_{0\le i\le j\le n}a^{(1)}_{(i,j)}y_iy_j+\sum_{0\le i\le n}b^{(1)}_iy_i+c^{(1)}\\
&\vdots\\
p^{(m)}(y_1,\ldots,y_n)&=\sum_{0\le i\leq j<n}a^{(m)}_{(i,j)}y_iy_j+\sum_{0\le i\le n}b^{(m)}_iy_i+c^{(m)},\\
\end{aligned}
\end{equation*}
where~$a^{(k)}_{(i,j)}$,~$b^{(k)}_i$, and~$c^{(k)}$ are elements in~$\bbF_q$ for every~$i$,$j$, and~$k$.

Equivalently, for convenience we view the system as one polynomial
$$
\bfp(\bfy)= \sum_{0<i\leq j<n}\bfa_{(i,j)}y_iy_j+\sum_{0<i<n}\bfb_iy_i+\bfc,
$$
where~$\bfa$,~$\bfb$, and~$\bfc$ are vectors in~$\bbF_q^m$. The goal is to find a solution~$\bfu =(u_1,\ldots,u_n)\in\bbF^n$ such that 
$\bfp(\bfu)=(0,\ldots,0)\in\bbF_q^m$.

Many MQ-based signature schemes were designed and analyzed, e.g., Unbalanced Oil and Vinegar~\cite{UOV}, Rainbow~\cite{Rainbow}, Gui~\cite{Gui} and more. In general, the public key of a MPKC is the set of coefficients of the quadratic polynomial system. A valid signature~$\bfs\in\bbF_q^n$ on a message~$\bfw\in\bbF_q^m$ is the solution to the quadratic system~$\bfp(\bfy)=\bfw$. Verifying the signature$\bfs$ against the message~$\bfw$ involves computing~$\bfw'=(p_1(\bfs),\ldots,p_m(\bfs))$, and the signature is accepted only when~$\bfw'=\bfw$. 

In addition to MQ-based signature schemes, hash functions based on multivariate polynomials of low degree have been discussed~\cite{MQHash} and analyzed~\cite{MQAnalysis}. Also, Applebaum \etal~\cite{LC} proves that under certain assumptions, there exists collision resistant hash functions which can be expressed as a multivariate polynomial of degree three over~$\bbF_2$, and yet it is unclear if this result extends to the larger fields employed here in. In the remainder of this paper we assume a polynomial hash function over~$\bbF_q$ of a constant degree~$d$.
%We are unsure this results would extend to larger finite fields, but assume a polynomial hash function over~$\bbF_q$ of a constant degree~$d$.

%In this paper, we replace the signature scheme (ECDSA) and hash function (SHA256 and RIPEMD-160) adopted in Bitcoin with MQ-based designs. Particularly, we formulate both signature verification and hash computation process as computing a set of multivariate polynomials of degree 3.

\subsection{Sharding and Coded Sharding}

Sharding-based blockchain systems partition the network into communities of constant size, allowing them to process and store disjoint sets of transactions in chain-like structures, namely \textit{shards}. Therefore, additional nodes help the system to process additional transactions, and scalability of throughput is achieved.

Conventional sharding schemes face two issues. First, mechanisms  such as random community reassignment are required for the security level to scale with network size. Second, the support for cross-shard transactions involves complicated communication and coordination protocols between communities. These added mechanisms hinder the improvement of performance.

The idea of \textit{coded sharding} was first introduced in Polyshard~\cite{Polyshard} to alleviate the security issue. In Polyshard, users are partitioned into communities. Every community~$k$ has an associated shard~$\bfY^{(t)}_k=(\bfY_k(1),\ldots,\bfY_k(t))$ defined as a chain of blocks, and each~$\bfY_k(\tau)$ stores verified transactions between members in community~$k$ in epoch~$\tau$. We emphasize that a user and a node are distinct entities; the former sends and receives coins, and the latter collects, verifies, and stores transactions.

Polyshard formulates the verification of a new block~$\bfX_k(t)$ against a shard~$\bfY_k^{(t-1)}$ at epoch~$t$  as computing a multivariate polynomial function, 
$$h^{(t)}_{k}=f^{(t)}(\bfX_k{(t)},\bfY^{(t-1)}_k),$$
where some of the function outputs affirm~$\bfX_k(t)$, and an indicator variable~$e^{(t)}_{k}\in \{0,1\}$ is generated accordingly. The verified block~$\bfY_k(t)=e^{(t)}_{k}\bfX_k(t)$ is then appended to the shard~$\bfY^{(t-1)}_{k}$, as~$e_k^{(t)}=1$ indicates the validity of~$\bfX_k(t)$.

Polyshard offers a novel separation between node and shard, as the actual verification is performed in a coded fashion, and a node does not verify or store transactions for any specific community. At the beginning of epoch~$t$, $K$~blocks~$\{\bfX_1(t),\bfX_2(t),...,\bfX_K(t)\}$ are generated, each stores transactions between members of a community. Meanwhile, every node~${i\in [N]}$ stores a \textit{coded shard}~${\widetilde{\bfY}_i^{(t-1)}=\sum^K_{k=1}{\ell_{i,k}} \bfY_k^{(t-1)}}$, where~${\bfl_i=(\ell_1,\ldots,\ell_K)\in\bbF_q^K}$ is the \textit{coding vector} of node~$i$ generated using a Lagrange polynomial.

Node~$i$ receives and linearly combines~$K$ blocks and produces a \textit{coded block}~$\widetilde{\bfX}_i(t)=\sum^K_{k=1}{\ell_{i,k}} \bfX_k(t)$. It then computes the verification function on coded data,
$$g^{(t)}_{i}=f^{(t)}(\widetilde{\bfX}_i(t),\widetilde{\bfY}^{(t-1)}_i),$$ 
and broadcasts the intermediate result~$g^{(t)}_{i}$ to all other nodes.

Given sufficiently many intermediate results from honest nodes, and a limited number of erroneous intermediate results from malicious nodes, a node can perform decoding and obtain the actual verification results~$\{h^{(t)}_1,\ldots,h^{(t)}_K\}$. Then, node~$i$ computes and linearly combines the verified blocks
to form the verified coded block~${\bfY_k(t)=\sum^K_{k=1}e^{(t)}_k{\ell_{i,k}} \bfX_k(t)}$  and appends it to the coded shard~$\widetilde{\bfY}_i^{(t-1)}$, which concludes the current epoch.

In general, Polyshard solves the blockchain impossibility triangle. First, additional computations of~$f^{(t)}$ allow the decoding of verification results for more shards, hereby supporting more transactions and scaling the throughput. Second, due to the underlying LCC mechanism, the number of malicious nodes that can be tolerated scales with the total number of nodes. Finally, since nodes perform verification on distinct linear combinations of all blocks, it follows that every node is partially processing every block, and hence decentralization is preserved. 

\section{Coded Verification}\label{section:Verification}

In this section, we provide a detailed description of our transaction verification scheme. First, we introduce our general settings and assumptions, as well as our 2-Dimensional sharding mechanism. Second, we formulate the verification of a new outgoing strip~$\bfh^{(t)}_{k}$ (to be defined shortly) against a shard~$\bfV^{(t-1)}$ in epoch~$t$ as as computing a polynomial function~$F^{(t)}$. Finally, we demonstrate the incorporation of this verification function~$F^{(t)}$ and Lagrange Coded Computing, showing how verification can be performed in a coded manner.
\subsection{Setting}
We distinguish between \textit{nodes} and \textit{users}. The~$N$ nodes are responsible for collecting, verifying and storing transactions; users create transactions and transfer coins between each other. 

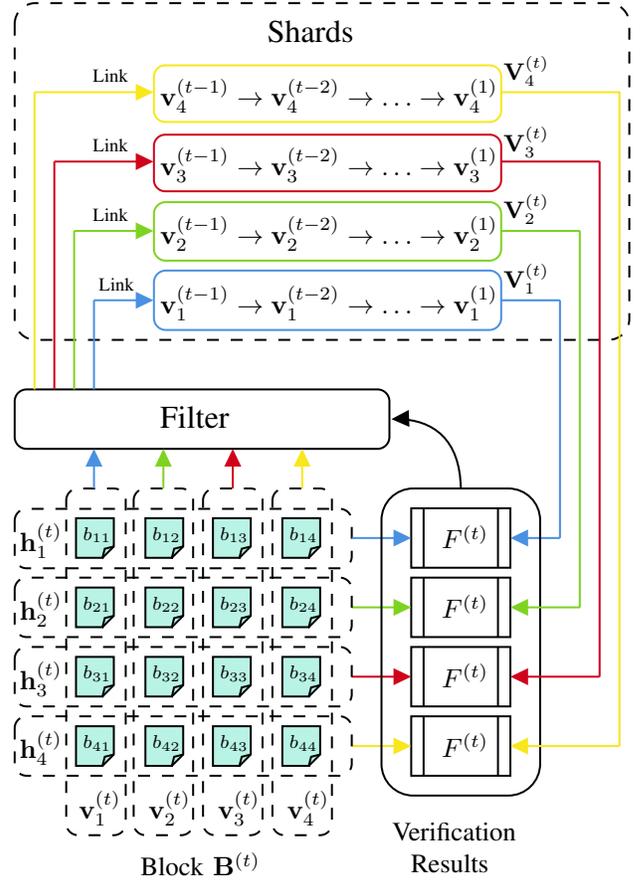
\begin{figure}
    \centering

\tikzset{every picture/.style={line width=0.75pt}} %set default line width to 0.75pt        

\begin{tikzpicture}[x=0.75pt,y=0.75pt,yscale=-1,xscale=1]
%uncomment if require: \path (0,851); %set diagram left start at 0, and has height of 851

%Rounded Rect [id:dp42945548680254353] 
\draw  [dash pattern={on 4.5pt off 4.5pt}] (145,30.56) .. controls (145,24.73) and (149.73,20) .. (155.56,20) -- (444.44,20) .. controls (450.27,20) and (455,24.73) .. (455,30.56) -- (455,179.44) .. controls (455,185.27) and (450.27,190) .. (444.44,190) -- (155.56,190) .. controls (149.73,190) and (145,185.27) .. (145,179.44) -- cycle ;
%Shape: Folded Corner [id:dp0003609511850086733] 
\draw  [fill={rgb, 255:red, 80; green, 227; blue, 194 }  ,fill opacity=0.4 ] (189.28,300) -- (175.93,300) -- (175.93,280) -- (195,280) -- (195,294.28) -- cycle -- (189.28,300) ; \draw   (195,294.28) -- (190.42,295.42) -- (189.28,300) ;
%Shape: Folded Corner [id:dp2919304842076309] 
\draw  [fill={rgb, 255:red, 80; green, 227; blue, 194 }  ,fill opacity=0.4 ] (189.28,335) -- (175.93,335) -- (175.93,315) -- (195,315) -- (195,329.28) -- cycle -- (189.28,335) ; \draw   (195,329.28) -- (190.42,330.42) -- (189.28,335) ;
%Shape: Folded Corner [id:dp23897682999166747] 
\draw  [fill={rgb, 255:red, 80; green, 227; blue, 194 }  ,fill opacity=0.4 ] (189.28,370) -- (175.93,370) -- (175.93,350) -- (195,350) -- (195,364.28) -- cycle -- (189.28,370) ; \draw   (195,364.28) -- (190.42,365.42) -- (189.28,370) ;
%Shape: Folded Corner [id:dp41094403867633233] 
\draw  [fill={rgb, 255:red, 80; green, 227; blue, 194 }  ,fill opacity=0.4 ] (189.28,405) -- (175.93,405) -- (175.93,385) -- (195,385) -- (195,399.28) -- cycle -- (189.28,405) ; \draw   (195,399.28) -- (190.42,400.42) -- (189.28,405) ;
%Shape: Folded Corner [id:dp07342852326224625] 
\draw  [fill={rgb, 255:red, 80; green, 227; blue, 194 }  ,fill opacity=0.4 ] (224.28,300) -- (210.93,300) -- (210.93,280) -- (230,280) -- (230,294.28) -- cycle -- (224.28,300) ; \draw   (230,294.28) -- (225.42,295.42) -- (224.28,300) ;
%Shape: Folded Corner [id:dp47166778112754004] 
\draw  [fill={rgb, 255:red, 80; green, 227; blue, 194 }  ,fill opacity=0.4 ] (224.28,335) -- (210.93,335) -- (210.93,315) -- (230,315) -- (230,329.28) -- cycle -- (224.28,335) ; \draw   (230,329.28) -- (225.42,330.42) -- (224.28,335) ;
%Shape: Folded Corner [id:dp974719235651776] 
\draw  [fill={rgb, 255:red, 80; green, 227; blue, 194 }  ,fill opacity=0.4 ] (224.28,370) -- (210.93,370) -- (210.93,350) -- (230,350) -- (230,364.28) -- cycle -- (224.28,370) ; \draw   (230,364.28) -- (225.42,365.42) -- (224.28,370) ;
%Shape: Folded Corner [id:dp27819153948760933] 
\draw  [fill={rgb, 255:red, 80; green, 227; blue, 194 }  ,fill opacity=0.4 ] (224.28,405) -- (210.93,405) -- (210.93,385) -- (230,385) -- (230,399.28) -- cycle -- (224.28,405) ; \draw   (230,399.28) -- (225.42,400.42) -- (224.28,405) ;
%Shape: Folded Corner [id:dp6877155609194154] 
\draw  [fill={rgb, 255:red, 80; green, 227; blue, 194 }  ,fill opacity=0.4 ] (258.35,300) -- (245,300) -- (245,280) -- (264.07,280) -- (264.07,294.28) -- cycle -- (258.35,300) ; \draw   (264.07,294.28) -- (259.5,295.42) -- (258.35,300) ;
%Shape: Folded Corner [id:dp6009562410441034] 
\draw  [fill={rgb, 255:red, 80; green, 227; blue, 194 }  ,fill opacity=0.4 ] (258.35,335) -- (245,335) -- (245,315) -- (264.07,315) -- (264.07,329.28) -- cycle -- (258.35,335) ; \draw   (264.07,329.28) -- (259.5,330.42) -- (258.35,335) ;
%Shape: Folded Corner [id:dp9940179398427438] 
\draw  [fill={rgb, 255:red, 80; green, 227; blue, 194 }  ,fill opacity=0.4 ] (258.35,370) -- (245,370) -- (245,350) -- (264.07,350) -- (264.07,364.28) -- cycle -- (258.35,370) ; \draw   (264.07,364.28) -- (259.5,365.42) -- (258.35,370) ;
%Shape: Folded Corner [id:dp15171840855423402] 
\draw  [fill={rgb, 255:red, 80; green, 227; blue, 194 }  ,fill opacity=0.4 ] (258.35,405) -- (245,405) -- (245,385) -- (264.07,385) -- (264.07,399.28) -- cycle -- (258.35,405) ; \draw   (264.07,399.28) -- (259.5,400.42) -- (258.35,405) ;
%Shape: Folded Corner [id:dp19521926778815146] 
\draw  [fill={rgb, 255:red, 80; green, 227; blue, 194 }  ,fill opacity=0.4 ] (293.35,300) -- (280,300) -- (280,280) -- (299.07,280) -- (299.07,294.28) -- cycle -- (293.35,300) ; \draw   (299.07,294.28) -- (294.5,295.42) -- (293.35,300) ;
%Shape: Folded Corner [id:dp26405874859431133] 
\draw  [fill={rgb, 255:red, 80; green, 227; blue, 194 }  ,fill opacity=0.4 ] (293.35,335) -- (280,335) -- (280,315) -- (299.07,315) -- (299.07,329.28) -- cycle -- (293.35,335) ; \draw   (299.07,329.28) -- (294.5,330.42) -- (293.35,335) ;
%Shape: Folded Corner [id:dp9701150107320393] 
\draw  [fill={rgb, 255:red, 80; green, 227; blue, 194 }  ,fill opacity=0.4 ] (293.35,370) -- (280,370) -- (280,350) -- (299.07,350) -- (299.07,364.28) -- cycle -- (293.35,370) ; \draw   (299.07,364.28) -- (294.5,365.42) -- (293.35,370) ;
%Shape: Folded Corner [id:dp17171520749608682] 
\draw  [fill={rgb, 255:red, 80; green, 227; blue, 194 }  ,fill opacity=0.4 ] (293.35,405) -- (280,405) -- (280,385) -- (299.07,385) -- (299.07,399.28) -- cycle -- (293.35,405) ; \draw   (299.07,399.28) -- (294.5,400.42) -- (293.35,405) ;
%Rounded Rect [id:dp7304394059000252] 
\draw  [dash pattern={on 4.5pt off 4.5pt}] (145,281.28) .. controls (145,277.96) and (147.69,275.28) .. (151,275.28) -- (309,275.28) .. controls (312.31,275.28) and (315,277.96) .. (315,281.28) -- (315,299.28) .. controls (315,302.59) and (312.31,305.28) .. (309,305.28) -- (151,305.28) .. controls (147.69,305.28) and (145,302.59) .. (145,299.28) -- cycle ;
%Rounded Rect [id:dp13320936440502384] 
\draw  [dash pattern={on 4.5pt off 4.5pt}] (171,271) .. controls (171,267.69) and (173.69,265) .. (177,265) -- (195,265) .. controls (198.31,265) and (201,267.69) .. (201,271) -- (201,434) .. controls (201,437.31) and (198.31,440) .. (195,440) -- (177,440) .. controls (173.69,440) and (171,437.31) .. (171,434) -- cycle ;
%Flowchart: Prodefined Process [id:dp14561628445088548] 
\draw   (345,275) -- (395,275) -- (395,305) -- (345,305) -- cycle ; \draw   (351.25,275) -- (351.25,305) ; \draw   (388.75,275) -- (388.75,305) ;
%Flowchart: Prodefined Process [id:dp5983447601167262] 
\draw   (345,310) -- (395,310) -- (395,340) -- (345,340) -- cycle ; \draw   (351.25,310) -- (351.25,340) ; \draw   (388.75,310) -- (388.75,340) ;
%Flowchart: Prodefined Process [id:dp4493691661128645] 
\draw   (345,345) -- (395,345) -- (395,375) -- (345,375) -- cycle ; \draw   (351.25,345) -- (351.25,375) ; \draw   (388.75,345) -- (388.75,375) ;
%Flowchart: Prodefined Process [id:dp6393419163787686] 
\draw   (345,380) -- (395,380) -- (395,410) -- (345,410) -- cycle ; \draw   (351.25,380) -- (351.25,410) ; \draw   (388.75,380) -- (388.75,410) ;
%Straight Lines [id:da4936889603275063] 
\draw [color={rgb, 255:red, 74; green, 144; blue, 226 }  ,draw opacity=1 ]   (420,290) -- (398,290) ;
\draw [shift={(395,290)}, rotate = 360] [fill={rgb, 255:red, 74; green, 144; blue, 226 }  ,fill opacity=1 ][line width=0.08]  [draw opacity=0] (8.93,-4.29) -- (0,0) -- (8.93,4.29) -- cycle    ;
%Straight Lines [id:da11780611544153152] 
\draw [color={rgb, 255:red, 74; green, 144; blue, 226 }  ,draw opacity=1 ][fill={rgb, 255:red, 74; green, 144; blue, 226 }  ,fill opacity=1 ]   (420,170) -- (420,290) ;
%Straight Lines [id:da45738864939938373] 
\draw [color={rgb, 255:red, 74; green, 144; blue, 226 }  ,draw opacity=1 ][fill={rgb, 255:red, 74; green, 144; blue, 226 }  ,fill opacity=1 ]   (420,170) -- (390,170) ;
%Straight Lines [id:da2536012891904744] 
\draw [color={rgb, 255:red, 74; green, 144; blue, 226 }  ,draw opacity=1 ]   (315,290) -- (342,290) ;
\draw [shift={(345,290)}, rotate = 180] [fill={rgb, 255:red, 74; green, 144; blue, 226 }  ,fill opacity=1 ][line width=0.08]  [draw opacity=0] (8.93,-4.29) -- (0,0) -- (8.93,4.29) -- cycle    ;
%Straight Lines [id:da6380453925086653] 
\draw [color={rgb, 255:red, 126; green, 211; blue, 33 }  ,draw opacity=1 ]   (315,325) -- (342,325) ;
\draw [shift={(345,325)}, rotate = 180] [fill={rgb, 255:red, 126; green, 211; blue, 33 }  ,fill opacity=1 ][line width=0.08]  [draw opacity=0] (8.93,-4.29) -- (0,0) -- (8.93,4.29) -- cycle    ;
%Straight Lines [id:da3734278233869084] 
\draw [color={rgb, 255:red, 208; green, 2; blue, 27 }  ,draw opacity=1 ]   (315,360) -- (342,360) ;
\draw [shift={(345,360)}, rotate = 180] [fill={rgb, 255:red, 208; green, 2; blue, 27 }  ,fill opacity=1 ][line width=0.08]  [draw opacity=0] (8.93,-4.29) -- (0,0) -- (8.93,4.29) -- cycle    ;
%Straight Lines [id:da9416303895535967] 
\draw [color={rgb, 255:red, 248; green, 231; blue, 28 }  ,draw opacity=1 ]   (315,395) -- (342,395) ;
\draw [shift={(345,395)}, rotate = 180] [fill={rgb, 255:red, 248; green, 231; blue, 28 }  ,fill opacity=1 ][line width=0.08]  [draw opacity=0] (8.93,-4.29) -- (0,0) -- (8.93,4.29) -- cycle    ;
%Rounded Rect [id:dp9918645829673649] 
\draw   (145,221) .. controls (145,217.69) and (147.69,215) .. (151,215) -- (328,215) .. controls (331.31,215) and (334,217.69) .. (334,221) -- (334,239) .. controls (334,242.31) and (331.31,245) .. (328,245) -- (151,245) .. controls (147.69,245) and (145,242.31) .. (145,239) -- cycle ;
%Rounded Rect [id:dp8507015526900763] 
\draw  [dash pattern={on 4.5pt off 4.5pt}] (205,271) .. controls (205,267.69) and (207.69,265) .. (211,265) -- (229,265) .. controls (232.31,265) and (235,267.69) .. (235,271) -- (235,434) .. controls (235,437.31) and (232.31,440) .. (229,440) -- (211,440) .. controls (207.69,440) and (205,437.31) .. (205,434) -- cycle ;
%Rounded Rect [id:dp39356915556822614] 
\draw  [dash pattern={on 4.5pt off 4.5pt}] (240,271) .. controls (240,267.69) and (242.69,265) .. (246,265) -- (264,265) .. controls (267.31,265) and (270,267.69) .. (270,271) -- (270,434) .. controls (270,437.31) and (267.31,440) .. (264,440) -- (246,440) .. controls (242.69,440) and (240,437.31) .. (240,434) -- cycle ;
%Rounded Rect [id:dp2091553267302202] 
\draw  [dash pattern={on 4.5pt off 4.5pt}] (275,271) .. controls (275,267.69) and (277.69,265) .. (281,265) -- (299,265) .. controls (302.31,265) and (305,267.69) .. (305,271) -- (305,434) .. controls (305,437.31) and (302.31,440) .. (299,440) -- (281,440) .. controls (277.69,440) and (275,437.31) .. (275,434) -- cycle ;
%Rounded Rect [id:dp4243644824120494] 
\draw   (330,281) .. controls (330,272.16) and (337.16,265) .. (346,265) -- (394,265) .. controls (402.84,265) and (410,272.16) .. (410,281) -- (410,404) .. controls (410,412.84) and (402.84,420) .. (394,420) -- (346,420) .. controls (337.16,420) and (330,412.84) .. (330,404) -- cycle ;
%Rounded Rect [id:dp07955613977877207] 
\draw  [color={rgb, 255:red, 74; green, 144; blue, 226 }  ,draw opacity=1 ] (215.23,161) .. controls (215.23,157.69) and (217.92,155) .. (221.23,155) -- (384.37,155) .. controls (387.68,155) and (390.37,157.69) .. (390.37,161) -- (390.37,179) .. controls (390.37,182.31) and (387.68,185) .. (384.37,185) -- (221.23,185) .. controls (217.92,185) and (215.23,182.31) .. (215.23,179) -- cycle ;
%Straight Lines [id:da1575791135352833] 
\draw [color={rgb, 255:red, 74; green, 144; blue, 226 }  ,draw opacity=1 ]   (185,265) -- (185,248) ;
\draw [shift={(185,245)}, rotate = 450] [fill={rgb, 255:red, 74; green, 144; blue, 226 }  ,fill opacity=1 ][line width=0.08]  [draw opacity=0] (8.93,-4.29) -- (0,0) -- (8.93,4.29) -- cycle    ;
%Straight Lines [id:da1681056582033882] 
\draw [color={rgb, 255:red, 126; green, 211; blue, 33 }  ,draw opacity=1 ]   (220,265) -- (220,248) ;
\draw [shift={(220,245)}, rotate = 450] [fill={rgb, 255:red, 126; green, 211; blue, 33 }  ,fill opacity=1 ][line width=0.08]  [draw opacity=0] (8.93,-4.29) -- (0,0) -- (8.93,4.29) -- cycle    ;
%Straight Lines [id:da34490018093803165] 
\draw [color={rgb, 255:red, 208; green, 2; blue, 27 }  ,draw opacity=1 ]   (255,265) -- (255,248) ;
\draw [shift={(255,245)}, rotate = 450] [fill={rgb, 255:red, 208; green, 2; blue, 27 }  ,fill opacity=1 ][line width=0.08]  [draw opacity=0] (8.93,-4.29) -- (0,0) -- (8.93,4.29) -- cycle    ;
%Straight Lines [id:da6691956573985254] 
\draw [color={rgb, 255:red, 248; green, 231; blue, 28 }  ,draw opacity=1 ]   (290,265) -- (290,248) ;
\draw [shift={(290,245)}, rotate = 450] [fill={rgb, 255:red, 248; green, 231; blue, 28 }  ,fill opacity=1 ][line width=0.08]  [draw opacity=0] (8.93,-4.29) -- (0,0) -- (8.93,4.29) -- cycle    ;
%Curve Lines [id:da5785747006641249] 
\draw    (370,265) .. controls (369.65,240.17) and (356.1,230.21) .. (337.91,229.96) ;
\draw [shift={(335,230)}, rotate = 357.66999999999996] [fill={rgb, 255:red, 0; green, 0; blue, 0 }  ][line width=0.08]  [draw opacity=0] (8.93,-4.29) -- (0,0) -- (8.93,4.29) -- cycle    ;
%Rounded Rect [id:dp5136107551402267] 
\draw  [color={rgb, 255:red, 126; green, 211; blue, 33 }  ,draw opacity=1 ] (215.23,126.38) .. controls (215.23,123.12) and (217.87,120.48) .. (221.13,120.48) -- (384.46,120.48) .. controls (387.73,120.48) and (390.37,123.12) .. (390.37,126.38) -- (390.37,144.1) .. controls (390.37,147.36) and (387.73,150) .. (384.46,150) -- (221.13,150) .. controls (217.87,150) and (215.23,147.36) .. (215.23,144.1) -- cycle ;
%Rounded Rect [id:dp6406231026980542] 
\draw  [color={rgb, 255:red, 208; green, 2; blue, 27 }  ,draw opacity=1 ] (215.23,92.14) .. controls (215.23,88.99) and (217.79,86.43) .. (220.94,86.43) -- (384.65,86.43) .. controls (387.81,86.43) and (390.37,88.99) .. (390.37,92.14) -- (390.37,109.29) .. controls (390.37,112.44) and (387.81,115) .. (384.65,115) -- (220.94,115) .. controls (217.79,115) and (215.23,112.44) .. (215.23,109.29) -- cycle ;
%Rounded Rect [id:dp2231860901765983] 
\draw  [color={rgb, 255:red, 248; green, 231; blue, 28 }  ,draw opacity=1 ] (215.23,57.14) .. controls (215.23,53.99) and (217.79,51.43) .. (220.94,51.43) -- (384.65,51.43) .. controls (387.81,51.43) and (390.37,53.99) .. (390.37,57.14) -- (390.37,74.29) .. controls (390.37,77.44) and (387.81,80) .. (384.65,80) -- (220.94,80) .. controls (217.79,80) and (215.23,77.44) .. (215.23,74.29) -- cycle ;
%Straight Lines [id:da8857333064159223] 
\draw [color={rgb, 255:red, 248; green, 231; blue, 28 }  ,draw opacity=1 ]   (155,65) -- (212,65) ;
\draw [shift={(215,65)}, rotate = 180] [fill={rgb, 255:red, 248; green, 231; blue, 28 }  ,fill opacity=1 ][line width=0.08]  [draw opacity=0] (8.93,-4.29) -- (0,0) -- (8.93,4.29) -- cycle    ;
%Straight Lines [id:da4911239989019678] 
\draw [color={rgb, 255:red, 248; green, 231; blue, 28 }  ,draw opacity=1 ]   (155,65) -- (155,215) ;

%Rounded Rect [id:dp5958819679514968] 
\draw  [dash pattern={on 4.5pt off 4.5pt}] (145,316) .. controls (145,312.69) and (147.69,310) .. (151,310) -- (309,310) .. controls (312.31,310) and (315,312.69) .. (315,316) -- (315,334) .. controls (315,337.31) and (312.31,340) .. (309,340) -- (151,340) .. controls (147.69,340) and (145,337.31) .. (145,334) -- cycle ;
%Rounded Rect [id:dp6650730982514046] 
\draw  [dash pattern={on 4.5pt off 4.5pt}] (145,351) .. controls (145,347.69) and (147.69,345) .. (151,345) -- (309,345) .. controls (312.31,345) and (315,347.69) .. (315,351) -- (315,369) .. controls (315,372.31) and (312.31,375) .. (309,375) -- (151,375) .. controls (147.69,375) and (145,372.31) .. (145,369) -- cycle ;
%Rounded Rect [id:dp8317479104785661] 
\draw  [dash pattern={on 4.5pt off 4.5pt}] (145,386) .. controls (145,382.69) and (147.69,380) .. (151,380) -- (309,380) .. controls (312.31,380) and (315,382.69) .. (315,386) -- (315,404) .. controls (315,407.31) and (312.31,410) .. (309,410) -- (151,410) .. controls (147.69,410) and (145,407.31) .. (145,404) -- cycle ;
%Straight Lines [id:da5556793887888882] 
\draw [color={rgb, 255:red, 208; green, 2; blue, 27 }  ,draw opacity=1 ]   (165,100) -- (212,100) ;
\draw [shift={(215,100)}, rotate = 180] [fill={rgb, 255:red, 208; green, 2; blue, 27 }  ,fill opacity=1 ][line width=0.08]  [draw opacity=0] (8.93,-4.29) -- (0,0) -- (8.93,4.29) -- cycle    ;
%Straight Lines [id:da08482574880318783] 
\draw [color={rgb, 255:red, 208; green, 2; blue, 27 }  ,draw opacity=1 ]   (165,100) -- (165,215) ;

%Straight Lines [id:da3182611744944901] 
\draw [color={rgb, 255:red, 126; green, 211; blue, 33 }  ,draw opacity=1 ]   (175,135) -- (212,135) ;
\draw [shift={(215,135)}, rotate = 180] [fill={rgb, 255:red, 126; green, 211; blue, 33 }  ,fill opacity=1 ][line width=0.08]  [draw opacity=0] (8.93,-4.29) -- (0,0) -- (8.93,4.29) -- cycle    ;
%Straight Lines [id:da057037022598003695] 
\draw [color={rgb, 255:red, 126; green, 211; blue, 33 }  ,draw opacity=1 ]   (175,135) -- (175,215) ;

%Straight Lines [id:da8642869496231522] 
\draw [color={rgb, 255:red, 74; green, 144; blue, 226 }  ,draw opacity=1 ]   (185,170) -- (212,170) ;
\draw [shift={(215,170)}, rotate = 180] [fill={rgb, 255:red, 74; green, 144; blue, 226 }  ,fill opacity=1 ][line width=0.08]  [draw opacity=0] (8.93,-4.29) -- (0,0) -- (8.93,4.29) -- cycle    ;
%Straight Lines [id:da48274328145868095] 
\draw [color={rgb, 255:red, 74; green, 144; blue, 226 }  ,draw opacity=1 ]   (185,170) -- (185,215) ;

%Straight Lines [id:da6321758702515445] 
\draw [color={rgb, 255:red, 126; green, 211; blue, 33 }  ,draw opacity=1 ]   (430,325) -- (398,325) ;
\draw [shift={(395,325)}, rotate = 360] [fill={rgb, 255:red, 126; green, 211; blue, 33 }  ,fill opacity=1 ][line width=0.08]  [draw opacity=0] (8.93,-4.29) -- (0,0) -- (8.93,4.29) -- cycle    ;
%Straight Lines [id:da6857412138503147] 
\draw [color={rgb, 255:red, 126; green, 211; blue, 33 }  ,draw opacity=1 ][fill={rgb, 255:red, 74; green, 144; blue, 226 }  ,fill opacity=1 ]   (430,135) -- (430,325) ;
%Straight Lines [id:da29153171413633205] 
\draw [color={rgb, 255:red, 126; green, 211; blue, 33 }  ,draw opacity=1 ][fill={rgb, 255:red, 74; green, 144; blue, 226 }  ,fill opacity=1 ]   (430,135) -- (390,135) ;
%Straight Lines [id:da45228999461011066] 
\draw [color={rgb, 255:red, 208; green, 2; blue, 27 }  ,draw opacity=1 ]   (440,360) -- (398,360) ;
\draw [shift={(395,360)}, rotate = 360] [fill={rgb, 255:red, 208; green, 2; blue, 27 }  ,fill opacity=1 ][line width=0.08]  [draw opacity=0] (8.93,-4.29) -- (0,0) -- (8.93,4.29) -- cycle    ;
%Straight Lines [id:da29778966589804123] 
\draw [color={rgb, 255:red, 208; green, 2; blue, 27 }  ,draw opacity=1 ][fill={rgb, 255:red, 74; green, 144; blue, 226 }  ,fill opacity=1 ]   (440,100) -- (440,360) ;
%Straight Lines [id:da9181788959529991] 
\draw [color={rgb, 255:red, 208; green, 2; blue, 27 }  ,draw opacity=1 ][fill={rgb, 255:red, 74; green, 144; blue, 226 }  ,fill opacity=1 ]   (440,100) -- (390,100) ;
%Straight Lines [id:da20605214720144427] 
\draw [color={rgb, 255:red, 248; green, 231; blue, 28 }  ,draw opacity=1 ]   (450,395) -- (398,395) ;
\draw [shift={(395,395)}, rotate = 360] [fill={rgb, 255:red, 248; green, 231; blue, 28 }  ,fill opacity=1 ][line width=0.08]  [draw opacity=0] (8.93,-4.29) -- (0,0) -- (8.93,4.29) -- cycle    ;
%Straight Lines [id:da29685941068492183] 
\draw [color={rgb, 255:red, 248; green, 231; blue, 28 }  ,draw opacity=1 ][fill={rgb, 255:red, 74; green, 144; blue, 226 }  ,fill opacity=1 ]   (450,65) -- (450,395) ;
%Straight Lines [id:da9083610287839969] 
\draw [color={rgb, 255:red, 248; green, 231; blue, 28 }  ,draw opacity=1 ][fill={rgb, 255:red, 74; green, 144; blue, 226 }  ,fill opacity=1 ]   (450,65) -- (390,65) ;

% Text Node
\draw (146,281) node [anchor=north west][inner sep=0.75pt]   [align=left] {$\bfh_1^{(t)}$};
% Text Node
\draw (359,282) node [anchor=north west][inner sep=0.75pt]   [align=left] {$F^{(t)}$};
% Text Node
\draw (359,317) node [anchor=north west][inner sep=0.75pt]   [align=left] {$F^{(t)}$};
% Text Node
\draw (359,352) node [anchor=north west][inner sep=0.75pt]   [align=left] {$F^{(t)}$};
% Text Node
\draw (359,387) node [anchor=north west][inner sep=0.75pt]   [align=left] {$F^{(t)}$};
% Text Node
\draw (176,416) node [anchor=north west][inner sep=0.75pt]   [align=left] {$\bfv_1^{(t)}$};
% Text Node
\draw (217,222) node [anchor=north west][inner sep=0.75pt]  [font=\large] [align=left] {Filter};
% Text Node
\draw (211,416) node [anchor=north west][inner sep=0.75pt]   [align=left] {$\bfv_2^{(t)}$};
% Text Node
\draw (246,416) node [anchor=north west][inner sep=0.75pt]   [align=left] {$\bfv_3^{(t)}$};
% Text Node
\draw (281,416) node [anchor=north west][inner sep=0.75pt]   [align=left] {$\bfv_4^{(t)}$};
% Text Node
\draw (217,161) node [anchor=north west][inner sep=0.75pt]   [align=left] {$\bfv_1^{(t-1)}\to \bfv_1^{(t-2) }\to\ldots\to \bfv_1^{(1)}$ };
% Text Node
\draw (217,126) node [anchor=north west][inner sep=0.75pt]   [align=left] {$\bfv_2^{(t-1)}\to \bfv_2^{(t-2) }\to\ldots\to \bfv_2^{(1)}$ };
% Text Node
\draw (217,91.5) node [anchor=north west][inner sep=0.75pt]   [align=left] {$\bfv_3^{(t-1)}\to \bfv_3^{(t-2) }\to\ldots\to \bfv_3^{(1)}$ };
% Text Node
\draw (217,56.5) node [anchor=north west][inner sep=0.75pt]   [align=left] {$\bfv_4^{(t-1)}\to \bfv_4^{(t-2) }\to\ldots\to \bfv_4^{(1)}$ };
% Text Node
\draw (146,315.72) node [anchor=north west][inner sep=0.75pt]   [align=left] {$\bfh_2^{(t)}$};
% Text Node
\draw (146,350.72) node [anchor=north west][inner sep=0.75pt]   [align=left] {$\bfh_3^{(t)}$};
% Text Node
\draw (146,385.72) node [anchor=north west][inner sep=0.75pt]   [align=left] {$\bfh_4^{(t)}$};
% Text Node
\draw (186,157) node [anchor=north west][inner sep=0.75pt]  [font=\scriptsize] [align=left] {Link};
% Text Node
\draw (271,27) node [anchor=north west][inner sep=0.75pt]  [font=\large] [align=left] {Shards};
% Text Node
\draw (334,432) node [anchor=north west][inner sep=0.75pt]   [align=left] {Verification\\ \ \ Results};
% Text Node
\draw (207,447) node [anchor=north west][inner sep=0.75pt]   [align=left] {Block $\bfB^{(t)}$};
% Text Node
\draw (177.93,283) node [anchor=north west][inner sep=0.75pt]  [font=\scriptsize] [align=left] {$b_{11}$};
% Text Node
\draw (212.93,283) node [anchor=north west][inner sep=0.75pt]  [font=\scriptsize] [align=left] {$b_{12}$};
% Text Node
\draw (247,283) node [anchor=north west][inner sep=0.75pt]  [font=\scriptsize] [align=left] {$b_{13}$};
% Text Node
\draw (282,283) node [anchor=north west][inner sep=0.75pt]  [font=\scriptsize] [align=left] {$b_{14}$};
% Text Node
\draw (177.93,318) node [anchor=north west][inner sep=0.75pt]  [font=\scriptsize] [align=left] {$b_{21}$};
% Text Node
\draw (212.93,318) node [anchor=north west][inner sep=0.75pt]  [font=\scriptsize] [align=left] {$b_{22}$};
% Text Node
\draw (247,318) node [anchor=north west][inner sep=0.75pt]  [font=\scriptsize] [align=left] {$b_{23}$};
% Text Node
\draw (282,318) node [anchor=north west][inner sep=0.75pt]  [font=\scriptsize] [align=left] {$b_{24}$};
% Text Node
\draw (177.93,353) node [anchor=north west][inner sep=0.75pt]  [font=\scriptsize] [align=left] {$b_{31}$};
% Text Node
\draw (212.93,353) node [anchor=north west][inner sep=0.75pt]  [font=\scriptsize] [align=left] {$b_{32}$};
% Text Node
\draw (247,353) node [anchor=north west][inner sep=0.75pt]  [font=\scriptsize] [align=left] {$b_{33}$};
% Text Node
\draw (282,353) node [anchor=north west][inner sep=0.75pt]  [font=\scriptsize] [align=left] {$b_{34}$};
% Text Node
\draw (177.93,388) node [anchor=north west][inner sep=0.75pt]  [font=\scriptsize] [align=left] {$b_{41}$};
% Text Node
\draw (212.93,388) node [anchor=north west][inner sep=0.75pt]  [font=\scriptsize] [align=left] {$b_{42}$};
% Text Node
\draw (247,388) node [anchor=north west][inner sep=0.75pt]  [font=\scriptsize] [align=left] {$b_{43}$};
% Text Node
\draw (282,388) node [anchor=north west][inner sep=0.75pt]  [font=\scriptsize] [align=left] {$b_{44}$};
% Text Node
\draw (390,150) node [anchor=north west][inner sep=0.75pt]  [font=\small] [align=left] {$\bfV_1^{(t)}$};
% Text Node
\draw (390,115) node [anchor=north west][inner sep=0.75pt]  [font=\small] [align=left] {$\bfV_2^{(t)}$};
% Text Node
\draw (390,81) node [anchor=north west][inner sep=0.75pt]  [font=\small] [align=left] {$\bfV_3^{(t)}$};
% Text Node
\draw (390,45) node [anchor=north west][inner sep=0.75pt]  [font=\small] [align=left] {$\bfV_4^{(t)}$};
% Text Node
\draw (183,122) node [anchor=north west][inner sep=0.75pt]  [font=\scriptsize] [align=left] {Link};
% Text Node
\draw (183,87) node [anchor=north west][inner sep=0.75pt]  [font=\scriptsize] [align=left] {Link};
% Text Node
\draw (183,52) node [anchor=north west][inner sep=0.75pt]  [font=\scriptsize] [align=left] {Link};

\end{tikzpicture}

\caption{Illustration of \textit{2-Dimensional Sharding} in a blockchain system with 4 shards. The block $B^{(t)}$ is horizontally sliced into \textit{outgoing strips}~$\bfh_1^{(t)},\bfh_2^{(t)},\bfh_3^{(t)},\bfh_4^{(t)}$ and vertically sliced into \textit{incoming strips}~$\bfv_1^{(t)},\bfv_2^{(t)},\bfv_3^{(t)},\bfv_4^{(t)}$. The outgoing strip $\bfh_k^{(t)}$ is verified against the corresponding shard~$\bfV_k^{(t)}$ in the verification function~$F^{(t)}$, for~$k=1,2,3,4$. Together, the verification results reveal the validity of every transaction in $\bfB^{(t)}$, and help to filter out the invalid transactions in the incoming strips. Finally, the filtered incoming strip $\bfv_k^{(t)}$ is linked to the corresponding shard~$\bfV_k^{(t)}$, for~$k=1,2,3,4$.}
    \label{fig:Fig.1}
\end{figure}

We partition users into~$K$ communities, each of a constant size which is independent of the total number of users. A \textit{shard}, or a sub-chain, is an append-only chain-like ledger containing UTXOs redeemable only to the members of an associated community.

Note that users are affiliated with communities, whereas nodes are not. Transactions are proposed and verified periodically during time intervals, called \textit{epochs}, denoted by a discrete time unit~$t$. In every epoch, every community has a constant number of active members, each of which proposes one transaction and transfers one coin to another arbitrary user.% where~$\mu\in [0,1]$ is referred as the \textit{activity factor} which is independent of the number of shards..

We formulate the block containing all transactions in epoch~$t$ as a matrix~$\bfB^{(t)}$,
\begin{equation}\label{eq: block}
\bfB^{(t)}=
\begin{bmatrix}
b_{1,1} & b_{1,2} & \ldots & b_{1,K} \\
b_{2,1} & b_{2,2} & \ldots & b_{2,K} \\
\vdots &\vdots&\ddots&\vdots\\
b_{K,1} & b_{K,2} & \ldots & b_{K,K} \\
\end{bmatrix},
\end{equation}
where every~$b_{k,r} \in \bbF_q^{Q\times R}$ is a \textit{tiny block}, formed as a concatenation of~$Q$ transactions  which redeem UTXOs in shard~$k$ and create new UTXOs in shard~$r$; each transaction~$\bfx\in\bbF_q^R$ is a vector of length~$R$ over some finite field~$\bbF_q$.

We partition the block $\bfB^{(t)}$ into \textit{outgoing strips} and \textit{incoming strips}, as shown in Fig.~\ref{fig:Fig.1}.
An {outgoing strip}~${\bfh_k^{(t)}=(b_{k,1},b_{k,2},\ldots,b_{k,K})\in (\bbF_q^{Q\times R})^K}$ is a vector containing all transactions in epoch~$t$ that redeem UTXOs in shard~$k$.  Similarly, an {incoming strip}~${\bfv_k^{(t)}=(b_{1,k},b_{2,k},\ldots,b_{K,k})\in (\bbF_q^{Q\times R})^K}$ stands for a collection of all transactions in epoch~$t$, which create new UTXOs only usable to members of community~$k$. 

In other words, every transaction in~$\bfh_k^{(t)}$ has a sender in community~$k$, and every transaction in~$\bfv_k^{(t)}$ has a receiver in community~$k$. Equivalently, one can view an outgoing strip~$\bfv^{(t)}_k$ as the~$k$'th row of matrix~$\bfB^{(t)}$, and incoming strip~$\bfv^{(t)}_k$ as the~$k$'th column of matrix~$\bfB^{(t)}$, i.e.,
\begin{equation*}
\bfB^{(t)}=\begin{bmatrix}(\bfv^{(t)}_{1})^\intercal,(\bfv^{(t)}_{2})^\intercal,\ldots,(\bfv^{(t)}_{K})^\intercal)\end{bmatrix}=\begin{bmatrix}\bfh^{(t)}_{1} \\ \bfh^{(t)}_{2}\\ \vdots \\ \bfh^{(t)}_{K}\end{bmatrix}.\\
\end{equation*} Moreover, a shard~$\bfV_k^{(t)}=\begin{pmatrix}\bfv_k^{(1)},\ldots,\bfv_k^{(t)}
\end{pmatrix}$ is a concatenation of \textit{incoming} strips associated with community~$k$, from epoch 1 to epoch~$t$. 

Note that we assume tiny blocks~$b_{k,r}$ to be of equal size. That is, the number of transactions at one epoch from shard~$k$ to shard~$r$ is identical, for every~$(k,r)\in [K]^2$. Moreover, in every epoch, we assume that each community has a constant portion of \textit{active members}, independent of the number of shards, and every active member initializes exactly one transaction which transfers one coin. Thus, the size of a strip is a constant, since it is dependent only on the portion of active members.
%Since the activity factor~$\mu$ is independent of the number of shards, it follows that an outgoing strip contains exactly~$\mu M$ transactions, and has a constant size of~$Q\cdot R\cdot K$.
%\CanranComment{How to say the size of an out going strip is also constant?}
\subsection{Polynomial Verification Function}

As mentioned in Section~\ref{section:background}.A, a new transaction is of the form~$\bfx_{new}=(\bfu_{new},\bfp_{new},\bfa_{new},\bfs_{new})$, where:
\begin{enumerate}
\item~$\bfu_{new}\in\bbF_{q^R}^{T^{(t-1)}\times 2}$ is a \textit{lookup matrix} used to index the previous transaction, where~$2^{T^{(t)}}$ is the number of transactions in the shard at epoch~$t$, and~$R$ is the length of a transaction.
\item~$\bfp_{new}\in\bbF_q^B$ is the sender's public key, containing all coefficients of an MQ system.
\item~$\bfa_{new}\in\bbF_q^C$ is the receiver's address, i.e., the hash value of the receiver's public key.
\item~$\bfs_{new}\in \bbF_q^D$ is the senders signature on~$\bfx_{new}'=(\bfu_{new},\bfp_{new},\bfa_{new})$
\end{enumerate}

Verifying~$\bfx_{new}$ includes two crucial parts:
\begin{enumerate}
    \item Fetching the corresponding old transaction~$\bfx_{old}=(\bfu_{old},\bfp_{old},\bfa_{old},\bfs_{old})$ from which~$\bfx_{new}$ redeems the~UTXO, and checking whether the hash value of~$\bfp_{new}$ matches~$\bfa_{old}$.
    \item Verifying that~$\bfs_{new}$ is a valid signature on the hash value of~$\bfx_{new}'$ by using the public key~$\bfp_{new}$.

\end{enumerate}

The lookup matrix~$\bfu_{new}$ has exactly one~1-entry and one~0-entry in each row. Hence, every transaction in~$\bfV^{(t)}$ can be uniquely indexed by a lookup matrix. The verifier views shard~$\bfV^{(t)}$ as~${T^{(t-1)}}$-dimensional tensor in~$(\bbF_{q^R})^{2\times 2\times ...\times 2}$, and therefore every transaction can be conveniently expressed as a tensor entry~$\bfV^{(t)}_{i_1,\ldots,i_{T^{(t-1)}}}\in \bbF_{q^R}$.

To fetch a transaction, one computes a multilinear polynomial,
\begin{equation*}
    \begin{split}
&f^{(t)}_{fetch}(\bfu,\bfV^{(t)})\\&= \sum_{(i_1,\ldots,i_{T^{(t-1)}})\in \{1,2\}^{T^{(t-1)}}} \left(\prod_{j=1}^{T^{(t-1)}} \bfu_{j,i_j}\right) \bfV^{(t)}_{i_1,\ldots,i_{T^{(t-1)}}}
    \end{split}
\end{equation*}
which takes a shard~$\bfV^{(t)}$ and a lookup table~$\bfu$ as inputs and yields the transaction~$\bfx_{\bfu}\in\bbF_{q^R}$ indexed by~$\bfu$. Note that the subscript~$k$ is omitted in~$f_{fetch}^{(t)}$ since the shard index is oblivious to the fetch function, i.e.,~$f_{fetch}^{(t)}$ can be applied to any shard. Since the fetch function sums up the product of~$T^{(t-1)}$ entries of~$\bfu$ and one entry of~$\bfV^{(t)}_k$, the degree of~$f^{(t)}_{fetch}$ is~$T^{(t-1)}+1$, which scales logarithmically with the number of transaction in a shard.

Based on Section 2.C, we assume a multivariate polynomial~$f_{hash1}:\bbF_q^B\to\bbF_q^C$ of degree~$d$ to serve as our first collision resistant hash function. Having obtained~$\bfx_{old}=f^{(t)}_{fetch}(\bfu_{new},\bfV^{(t)}_k)$, the verifier then checks whether~$f_{hash1}(\bfp_{new})=\bfa_{old}$ holds. The address check function is expressed as a polynomial,
$$f_{checkAddr}(\bfp,\bfa)=f_{hash1}(\bfp)-\bfa.$$

Since the degree of~$f_{hash1}$ is~$d$, and~$\bfa_{new}$ is the output of polynomial~$f_{fetch}^{(t)}$ with degree~$T^{(t-1)}+1$, it follows that~$f_{checkAddr}$ has a degree of~$\max(T^{(t-1)}+1,d)$.
Note that~$\bfp_{new}$ is accepted when~$\bfr_1=f_{checkAddr}(\bfp_{new},\bfa_{old})\in\bbF_q^C$ is the all-zero vector. This concludes up part 1.

In part 2, the verifier checks the validity of the signature~$\bfs_{new}$. She first computes a hash digest~$\bfw=f_{hash2}(\bfu_{new}, \bfp_{new},\bfa_{new})=(w_1,\ldots,w_E)\in \bbF_q^E$, where~$f_{hash2}: \bbF_q^{A+B+C}\to\bbF_q^E$ is our second collusion resistant hash function of degree~$d$. Later, the verifier checks whether~$f_{MQ}(\bfp_{new},\bfs_{new})=\bfw$ holds, where,
$$
f_{MQ}(\bfp,\bfs)= \sum_{0<i\leq j<n}\bfa_{(i,j)}s_is_j+\sum_{0<i<n}\bfb_is_i+\bfc,
$$
and~$\bfa, \bfb, \bfc \in \bbF_q^E$ are vectors stored in~$\bfp_{new}$, serving as coefficients of the MQ problem. Equivalently, the verification of a signature~$\bfs$ in a transaction~$\bfx=(\bfu,\bfp,\bfa,\bfs)$ can be expressed as a polynomial,
$$
f_{checkSig}(\bfx)= f_{MQ}(\bfp,\bfs)-f_{hash2}(\bfu,\bfp,\bfa),
$$
whose degree is~$d$, since~$\deg f_{MQ}=\deg f_{hash2}=d$. Note that~$\bfs_{new}$ is accepted only when~${\bfr_2=f_{checkSig}(\bfx_{new})=0}$. This concludes part 2.

%Let~$\bfr_1,\bfr_2$ be the results of part 1 and part 2, respectively. We consider their concatenation~$\bfr$ as the result of computing a multivariate polynomial~$f^{(t)}(\bfx,\bfV^{(t)})$ of degree~$\max(T^{(t-1)}+1,d)$. 
Let~$\bfr$ be the concatenation of~$\bfr_1,\bfr_2$ (the results of part 1 and part 2), which is seen as the result of computing a multivariate polynomial~$f^{(t)}(\bfx,\bfV^{(t)})$ of degree~$\max(T^{(t-1)}+1,d)$.
The transaction~$\bfx$ is accepted if and only if~$\bfr=f^{(t)}(\bfx,\bfV^{(t)})=0$.
\begin{remark}
We consider a realistic blockchain systems in which~${T^{(t-1)}\ge d}$. Therefore,~$\deg f^{(t)}=T^{(t-1)}+1$, where~$2^{T^{(t-1)}}$ is the number of transaction in a shard a epoch~$t$. That is, the degree of~$f^{(t)}$ scales logarithmically with the number of transaction in a shard. 
\end{remark}

Moreover, since the coins redeemed by transactions in~$\bfh^{(t)}_{k}=(\bfx_1,\ldots,\bfx_{QK})$ all reside in~$\bfV^{(t)}_{k}$, we define a multivariate polynomial,
\begin{align*}
	F^{(t)}&(\bfh^{(t)},\bfV^{(t-1)})=\\
	&(f^{(t)}(\bfx_1,\bfV^{(t-1)}),\ldots,f^{(t)}(\bfx_{QK},\bfV^{(t-1)})),	
\end{align*}
which yields a result matrix~$\bfR_k^{(t)}\in\bbF_q^{(C+E)\times QK}$, whose every row is the result of computing~$f^{(t)}$ on~$\bfV^{(t)}_k$ and one transaction in~$\bfh^{(t)}_{k}$. 

Based on the result matrix, an \textit{indicator vector}~$\bfe^{(t)}_k=(e^{(t)}_{k,1},\ldots,e^{(t)}_{k,QK})\in \{0,1\}^{QK}$ is generated accordingly. For all~$j\in[QK]$,~$e_j^{(t)}$ indicates the validity of the respective transaction, i.e.,~$e^{(t)}_j=1$ when~$\bfr_j=f^{(t)}(\bfx_j,V^{(t)})=0$, otherwise ~$e^{(t)}_j=0$. As shown in Fig.~\ref{fig:Fig.1}, the indicator vectors $\bfe^{(t)}_1,\bfe^{(t)}_2,\ldots,\bfe^{(t)}_K$ reveal the validity of every transaction in the block $\bfB^{(t)}$, and are used to filter out the invalid transactions in the incoming strips before they are being appended to the corresponding shards. 

\subsection{Coded Computation}
Now that the verification of outgoing strips has been formulated as a low degree polynomial, we turn to describe how it is conducted in a coded fashion. In detail, every shard~$k\in[K]$ is assigned a unique scalar~$\omega_k\in\bbF_q$, and every node~$i\in[N]$ is assigned a unique scalar~$\alpha_i\in\bbF_q$. Node~$i$ computes a node-specific \textit{coding vector} using~$\{\omega_k\}_{k=1}^K$ and~$\alpha_i$, 
$$\bfl_i=(\ell_1(\alpha_i),\ell_2(\alpha_i),...,\ell_K(\alpha_i))\in\bbF_q^K,$$
where
$$\ell_k(z)=\prod_{j\neq k} \frac{z-\omega_j}{\omega_k - \omega_j}.$$

A coded outgoing strip for node~$i$ is defined as the linear combination of all outgoing strips,
$$ 
\widetilde{\bfh}^{(t)}_{i}=\bfl_i\begin{bmatrix}\bfh^{(t)}_1\\
\vdots\\\bfh^{(t)}_K\\\end{bmatrix}=\bfl_i\bfB^{(t)},
$$
and for the coded incoming strip,
$$
\widetilde{\bfv}^{(t)}_{i}=\bfl_i\begin{bmatrix}\bfv^{(t)}_1\\
\vdots\\\bfv^{(t)}_K\\\end{bmatrix}=\bfl_i(\bfB^{(t)})^\intercal.
$$

Equivalently, one can think of~$\widetilde{\bfh}^{(t)}_{i}$ and~$\widetilde{\bfv}^{(t)}_{i}$ as evaluations of Lagrange polynomials~$\psi^{(t)}(z)$ and~$\phi^{(t)}(z)$ at~$\alpha_i$, respectively, where
$$\psi^{(t)}(z)=\sum_{k=1}^K\bfh^{(t)}_{k}\prod_{j\neq k} \frac{z-\omega_j}{\omega_k - \omega_j},$$
and,
$$
\phi^{(t)}(z)=\sum_{k=1}^K\bfv^{(t)}_{k}\prod_{j\neq k} \frac{z-\omega_j}{\omega_k - \omega_j}.
$$
Note that both~$\psi^{(t)}$ and~$\phi^{(t)}$ are of degree~$K-1$, also note that the outgoing strip~$\bfh^{(t)}_{k}$ equals to~$\psi^{(t)}(\omega_k)$, and the incoming strip~$\bfv^{(t)}_{k}$ equals to~$\phi^{(t)}(\omega_k)$, for all~$k=1,2,\dots,K$. 

Every node~$i$ stores a coded shard~$\widetilde{\bfV}^{(t)}_{i}$, defined as a node-specific linear combination of all shards,
$$\widetilde{\bfV}^{(t)}_{i}=\bfl_i\begin{pmatrix}{\bfV}^{(t)}_1\\
\vdots\\{\bfV}^{(t)}_K\end{pmatrix}=(\phi^{(1)}(\alpha_i),\ldots,\phi^{(t)}(\alpha_i)),$$

In the epoch~$t$, node~$i$ receives a coded outgoing strip~$\widetilde{\bfh}^{(t)}_{i}$ and coded incoming strip~$\widetilde{\bfv}^{(t)}_{i}$ designed for it (the detailed scheme is discussed in Section~\ref{section:propagation}). Every node computes the polynomial verification function~$F^{(t)}$ on~$\widetilde{\bfh}^{(t)}_{i}$ and the locally stored~$\widetilde{\bfV}_t^{i}$, and obtain a coded result matrix, 
$$\widetilde{\bfR}^{(t)}_{i}=F^{(t)}(\widetilde{\bfh}^{(t)}_{i}, \widetilde{\bfV}^{(t)}_{i}).$$

The resulting coded indicators can be seen as the evaluation of the following polynomial at~$\alpha_i$,
$$
\widetilde{\bfR}^{(t)}_{i}=F^{(t)}(\psi^{(t)}(\alpha_i),(\phi^{(1)}(\alpha_i),\ldots,\phi^{(t)}(\alpha_i))).
$$

After receiving the result vectors from sufficiently many nodes, or equivalently, sufficiently many distinct evaluations of the function~$F^{(t)}(\psi^{(t)}(z),\phi^{(t-1)}(z),\ldots,\phi^{(1)}(z))$, node~$i$ can obtain the coefficients of~$F^{(t)}$ as a polynomial in~$z$; this is achieved by decoding the Lagrange code. 

By evaluating~$F^{(t)}$ on~$\omega_1,\omega_2,\ldots,\omega_K$, node~$i$ obtains the uncoded result matrices~$\bfR_1^{(t)},\bfR_2^{(t)},\ldots,\bfR_K^{(t)}$, from which the validity of every transaction in epoch~$t$ can be determined. These verification results will be reflected in the appended incoming strips, described next.
\subsection{Coded Appending}

Following the coded verification, nodes append the coded \textit{incoming} strip to their coded shard, after setting to zero the parts of it which failed the verification process. This is done as follows.

Let~$\bfx_{s,k,r}$ be the~$s$'th transaction in~$b^{(t)}_{k,r}$, where~${s\in[Q]}$ and~${ k,r\in[K]}$, which is invalid if the corresponding indicator variable~$e^{(t)}_{k,(r-1)K+s}$ is zero. Recall that~$b^{(t)}_{k,r}\in {\bbF_q^{Q\times R}}$ is a tiny block which contains~$Q$ transactions which redeem coins in shard~$k$ and create coins in shard~$r$ at epoch~$t$. Therefore,~$K$ transactions are abandoned as a result of one transaction being invalid.

In a coded incoming strip, an invalid transaction~$\bfx_{s,k,r}$ is linearly combined with other transactions in~$\{\bfx_{s,k,1},\ldots,\bfx_{s,k,K}\}$ from other columns of~$\bfB^{(t)}$, forming~$\widetilde{\bfx}_{s,k,i}$, as the coded~${((k-1)K+s)}$'th transaction (i.e., the~$s$'th coded transaction in the~$k$'th coded tiny block), in the coded incoming strip~$\widetilde{\bfv}_i^{(t)}$. If a transaction~$\bfx_{s,k,r}$ fails verification for some~$r\in[K]$, i.e., if the respective~$e^{(t)}_{k,(r-1)K+s}$ is zero, then node~$i$ replaces~$\widetilde{\bfx}_{s,k,i}$ with zeros, and appends the resulting coded incoming strip~$\widetilde{\bfv}^{(t)}_{i}$ to its locally stored coded shard~$\widetilde{\bfV}^{(t)}_i$.

\begin{remark}
We define the Collateral Invalidation (CI) rate as the number of transactions that are abandoned due to one invalid transaction, normalized by the total number of transactions processed in one epoch.

In Polyshard, one invalid transaction causes an entire block to be abandoned. Thus, Polyshard has an CI rate of~$\frac{1}{K}$. In our scheme, when a transaction is invalid, we erase one coded transaction in every coded incoming strip, which is equivalent to wipe out K transactions. In total,~$QK^2$ transactions are verified in one epoch, resulting in an CI rate of~$\frac{1}{KQ}$, which is~$Q$ times smaller (i.e., better) than Polyshard. 
\end{remark}

\section{Coded Propagation}\label{section:propagation}

As discussed in the previous section, node~$i$ performs coded verification and coded appending after receiving both coded incoming~$\widetilde{\bfh}_i^{(t)}$ and coded outgoing strips~$\widetilde{\bfv}_i^{(t)}$. We call such a node \textit{complete}, and otherwise, it is \textit{incomplete}. 

In this section, we present a time-efficient protocol in which nodes communicate in order to turn all nodes to complete nodes. We assume that, in every epoch,~$K$ honest leader nodes are randomly selected, each of which collects transactions that redeem UTXOs from shard~$k$ and forms an outgoing strip~$\bfh^{(t)}_{k}$, for all~$k\in [K]$. Without lost of generality, we let the node with index~$k$ be the leader of shard~$k$. Note that the leader selection mechanism is outside the scope of this paper.

We assume that the communication is synchronous, i.e., happens in \textit{rounds} (not to be confused with epochs). To account for limited bandwidth in real-world systems, we define the \textit{capacity} of a node to be the amount of data it can send and receive during one round. The capacity is normalized by the size of a strip, which is constant and independent of the number of nodes or shards.

We split the propagation into three stages. At the first stage, the~$K$ leader nodes communicate in order to obtain all coded outgoing strips. At the second stage they communicate to obtain all coded incoming strips. At the third stage, the~$K$ leader nodes initiate a completion process for the remaining~$N-K$ nodes. In total,
$${\frac{2(\sqrt{K}-1)}{D}+\log_{(D+1)}\frac{N}{K}}+1$$ rounds are required to deliver all coded strips to every node, where~$D$ is the capacity of each node.

\subsection{Stage One}
In the first stage, the leader nodes acquire their coded outgoing strips~$\{\widetilde{\bfh}_{k}^{(t)}\}_{k=1}^K$ by the following communication protocol, which consists of the \textit{preparation} phase and the \textit{shooting} phase.

First, we assign the~$K$ leader nodes into~$m=\sqrt{K}$ groups, each of size~$m$. We let~$(x,y)$ to represent the~$y$'th node in the~$x$'th group, and let~${\bfh}_{(x,y)}$ be its outgoing strip, for~$x,y \in [m]$.

In the preparation phase, the outgoing strips~${\{\bfh_{(x,1)}^{(t)},\ldots,\bfh_{(x,m)}^{(t)}\}}$ are rotated internally in every group~$x\in[m]$, until they are shared by every group member. In detail, in the~$n$'th round, node~${(x,y)}$ sends the outgoing strip~$\bfh_{(x,y)}^{(t)}$ to node~$(x,y+\delta(n,d))$, and receives the outgoing strip~$\bfh^{(t)}_{{(x,y-\delta(n,d}))}$ from node~$({x,y-\delta(n,d)})$, for all~$d\in [D]$, where
$$
\delta(n,d)=(n-1)D+d.
$$
Clearly,~$\frac{m-1}{d}$ rounds are required for every node~$({x,y})$ to obtain outgoing strips~${\{\bfh_{(x,1)}^{(t)},\ldots,\bfh_{(x,m)}^{(t)}\}}$.

Then, in the shooting phase, all nodes in the set~$\{(1,y),\ldots,(m,y)\}$ linearly combine the strips received in the preparation phase, and forward to each other, for every~$y\in[m]$. Formally, in round~$n$, node~$({x,y})$ sends the packet
$$
\sum_{u=1}^m \ell_{(x+\delta(n,d),y),(x,u)}\bfh^{(t)}_{(x,u)},
$$
to node~$(x+\delta(n,d),y)$, and receives the packet,
$$
\sum_{u=1}^m \ell_{(x,y),(x-\delta(n,d),u)}\bfh^{(t)}_{(x-\delta(n,d)),u)},
$$
from node~$(x-\delta(n,d),y)$, for every~$d\in[D]$. Recall that~$\ell_{i,k}$ stands for the coding coefficient for strip~$\bfh^{(t)}_k$ or~$\bfv^{(t)}_k$ of node~$i$.

Recall that, 
$$\widetilde{\bfh}_{(x,y)}^{(t)}=\sum_{u,v\in[m]}\ell_{(x,y),(u,v)}\bfh^{(t)}_{(u,v)},$$
and therefore, node~$(x,y)$ can obtain its coded outgoing strip~$\widetilde{\bfh}^{(t)}_{(x,y)}$ by summing up the packets received in the~$\frac{m-1}{d}$ rounds of the shooting phase, together with its own packet,
$$\sum_{u=1}^m \ell_{(x,y),(x,u)}\bfh^{(t)}_{(x,u)}.$$

Overall, this stage requires~$\frac{2(m-1)}{D}$ rounds, and each leader node receives~$\sqrt{K}-1$ strips.

\subsection{Stage Two}
In the second stage, having obtained the coded outgoing strips, the leader nodes exchange data to acquire their coded incoming strips, and hereby be complete.

Node~$k$ computes and forwards a \textit{data drop}~$w^{(t)}_{k,r}$ to node~$r$, for all~$r\in [K]$ including itself, where
$$
w^{(t)}_{k,r}={\bfh}^{(t)}_{k}(\bfl_{r})^\intercal=\bfl_{k}({\bfB^{(t)}})^\intercal(\bfl_{r})^\intercal,
$$ with~$\bfB^{(t)}$ as in~\eqref{eq: block}.

Meanwhile, node~$k$ receives~$K$ drops~$w^{(t)}_{1,k},...,w^{(t)}_{K,k}$, from each leader node.
Note that this operation can be finished in one round, as the size of one drop is~$\frac{1}{K}$ of a strip (i.e., the size of one tiny block) and we assume~$D\geq 1$.

Then, node~$k$ aligns the received data drops into a vector,
\begin{equation}\label{eq:drops}
\bfw^{(t)}_{leaders,k}=(w^{(t)}_{1,k},...,w^{(t)}_{K,k})=\bfL_{leaders}({\bfB^{(t)}})^\intercal(\bfl_k)^\intercal,
\end{equation}
where~$\bfL_{leaders}\in\bbF_q^{K\times K}$ is a matrix made of coding vectors from all~$K$ leader nodes, i.e.,
$$
\bfL_{leaders}=\begin{bmatrix}
\bfl_{1} \\ \bfl_{2}\\ \vdots \\ \bfl_{K}
\end{bmatrix}.
$$
Node~$k$ then obtains its coded \textit{incoming} strip~$\widetilde{\bfv}^{(t)}_{r}$ by multiplying \eqref{eq:drops} from the left by~$\bfL_{leaders}^{-1}$ (which exists according to the MDS property of the Lagrange matrix, see Section~\ref{section:background}),
\begin{equation*}
    \begin{split}
        (\bfL_{leaders})^{-1}\bfw^{(t)}_{{leaders},k}&=(\bfL_{leaders})^{-1}\bfL_{leaders}({\bfB^{(t)}})^\intercal(\bfl_k)^\intercal\\
        &=({\bfB^{(t)}})^\intercal(\bfl_k)^\intercal\\
        &=(\bfl_k\bfB^{(t)})^\intercal\\
        &=(\widetilde{\bfv}^{(t)}_{k})^\intercal.
    \end{split}
\end{equation*}

Having obtained the coded incoming and outgoing strips, the~$K$ leader nodes are now complete. This marks the end of stage two, which takes only one round, and each node receives data of of size 1 strip.
\begin{remark}
Inspired by Product-Matrix MBR codes~\cite{ProductMatrix}, our scheme achieves the minimal bandwidth required to complete a node, i.e., the size of two strips.

In particular, our scheme allows a node to obtain its coded incoming strip by downloading~$K$ data drops, each computed from a distinct coded outgoing strip. Similarly, a node can obtain its coded outgoing strip by downloading~$K$ data drops, each computed from a distinct coded incoming strip. 
\end{remark}
\subsection{Stage Three}
In this final stage, we employ the~$K$ complete leader nodes to complete the other~$N-K$ nodes, hereby finalizing our scheme.

As discussed in Remark 3, an incomplete node can contact~$K$ complete nodes, download~$2$ data drops computed from the incoming and outgoing strips from each one, and complete itself.

We observe that every~$K$ complete nodes together can complete~$KD$ incomplete nodes in two rounds, and recall that the size of a data drop is~$\frac{1}{K}$ of the size of a strip (e.g., for~$D=1$ a node can forward the equivalent of one strip every round). Also, recall that the capacity~$D$ is the amount of data a node can send and receive in one round, normalized by the size of a strip. Hence, a complete node can forward one data drop to each of the~$KD$ incomplete nodes in one round. 

Further,~$K$ complete nodes can guarantee that each of~${KD}$ incomplete nodes receive~$K$ different data drops in one round, which are sufficient for each of them to obtain one of its coded strips. Therefore, it requires \textit{two} rounds to complete~$KD$ nodes, one for the incoming and for for the outgoing strips.

Let~$a_n$ be the number of nodes completed in round~$2n$, and let~$S_n=\sum_{i=0}^na_i$. Since~$K$ leader nodes have been completed in the previous stage, we have~$a_0=K$. In each two-round pair, we assign~$S_{n-1}$ (for~$n\geq 1$) completed nodes into groups of size~$K$, and let each group complete a set of~$KD$ nodes, where different sets are disjoint. According to the above, by the end of the second round we have~$a_1=KD$, and hence~${S_1=a_0+a_1=K+KD}$.

It is readily verified that~$K\mid S_n$ is true for all~$n\geq 0$. Hence, at the beginning of round~$2n$, there exists~$\frac{S_{n-1}}{K}$ groups of completed nodes, each of size~$K$, including the group of leader nodes.  We obtain the following expression for the number of nodes completed in round~$2n$,
$$
a_n=\frac{S_{n-1}}{K}\cdot KD=S_{n-1}D.
$$
Since~$a_n=S_n-S_{n-1}$, it follows that
$$
S_n-S_{n-1}=S_{n-1}D,\text{ and hence } S_n=S_{n-1}(D+1).
$$
Finally, we deduce the expression for the total number of completed nodes at round~$2n$,
$$
S_n = K(D+1)^n.
$$
Letting~$S_n=N$, we find that~$n=\log_{(D+1)}\frac{N}{K}$. Therefore, it takes~$2\log_{(D+1)}\frac{N}{K}$ rounds to finish the third stage, and hereby finish the propagation of coded strips.

\section{Discussion}\label{section:discussion}
In this section, we discuss the key improvements of our scheme.

\subsection{Support for Cross-Shard Transactions}
We propose 2-Dimensional sharding, which provides inherent support for cross-shard transactions by partitioning the transactions by sender and receiver. The novelty lies on the fact that we decouple the verification and the appending of transactions, i.e., the outgoing strip is verified, while the incoming strip is appended. 

Meanwhile, our scheme allows the verification results to be reflected in the appended transactions. That is, by examining the verification results of outgoing strips, a node can remove the invalid coded transactions in the coded outgoing strip, before linking it to the locally stored chain.

Finally, compared with Polyshard, our scheme induces a relatively low CI rate of~$\frac{1}{KQ}$, which is~$Q$ times better in comparison with Polyshard. That is, the number of valid transaction abandoned due to an invalid transaction is~$Q$ times smaller than Polyshard.

\subsection{Security}
We discuss the factors that affects the security level in our scheme, as well as countermeasures. We formulated the verification of a new strip against a shard as computing a multivariate polynomial function~$F^{(t)}(\bfh^{(t)},\bfV^{(t)})$, and incorporate it with Lagrange Coded Computing. The degree of the~$F$ is~${T^{(t-1)}+1}$ in the elements of~$\bfh^{(t)}$ and~$\bfV^{(t)}$, where~$M^{(t-1)}=2^{T^{(t-1)}}$ is the number of transaction in a shard at the beginning of epoch~$t$. 

By the properties of Lagrange Coded Computing, to guarantee correct computation, it is required that 
\begin{equation}\label{eq:limit}
\begin{split}
    (\log_2M^{(t-1)}+1)(K-1)+S+2A+1&\leq N,
\end{split}
\end{equation}
where~$A$ represents the number of malicious nodes and~$S$ is the number of stragglers.% (i.e., nodes whose results are not received by node~$i$).

To guarantee security, we must assure a correct verification result for every individual node~$i$, even in the extreme case where the entire set of~$A$ malicious nodes in the system is included in the~$N-S$ nodes from which node~$i$ receives intermediate results. Therefore, every node should receive intermediate results from 
\begin{equation}\label{eq: limit2}
\begin{split}
N-S\geq(\log_2M^{(t-1)}+1)(K-1)+2A+1.
\end{split}
\end{equation}nodes before performing decoding, 

We hereby define~$\beta=\frac{A}{N}$ as the ratio of malicious nodes in our system, whose maximum value that can be tolerated serves as the measure for security. Also, as it is infeasible for every node to wait for every other node, we require nodes to perform decoding once after receiving intermediates results from a certain portion of all nodes, defined as~$\gamma=1-\frac{S}{N}$. By~\eqref{eq: limit2}, we derive the requirement for~$\beta$ and~$\gamma$:
\begin{equation}\label{eq: limit3}
\begin{split}
\gamma-2\beta \geq \frac{(\log_2M^{(t-1)}+1)(K-1)+1}{N}.
\end{split}
\end{equation}
For large~$N, M$ and~$K$, the right side of \eqref{eq: limit3} is approximately 
%$$\log_2M^{(t-1)}\frac{K}{N},$$
$$\frac{K}{N}\log_2M^{(t-1)},$$
and hence \eqref{eq: limit3} becomes
%\begin{equation}\label{eq: limit4}
%\begin{split}
%2\beta \leq \gamma-  \log_2M^{(t-1)}\frac{K}{N}.
%\end{split}
%\end{equation}
\begin{equation}\label{eq: limit4}
	\begin{split}
		2\beta \leq \gamma - \frac{K}{N}\log_2M^{(t-1)}.
	\end{split}
\end{equation}
To maintain a consistent security level (i.e., to stabilize the upper bound of~$\beta$), the right side of \eqref{eq: limit4} should stay constant. 

First, as more users join the system, new communities are formed, which leads to an increase of~$K$. In this case, the number of nodes~$N$ should scale with~$K$ to counter the drop of tolerable number of malicious nodes.

Second,~$N$ should also scale logarithmically with the number of transactions in one shard, which grows with~$t$ since new transactions are added to each shard in every epoch. 

To show the applicability of our scheme, we examine the parameters of a real-world blockchain system similar to Ethereum 2.0~\cite{ETC} with~$64$ shard,~$10000$ nodes and assume a billion ($M\approx 2^{30}$) transactions in each shard. In such a system,~$\beta$ and~$\gamma$ must satisfy  
$$
\gamma- 2\beta \geq  0.192.
$$

To maximize the number of tolerable malicious nodes, honest nodes should performs decoding after receiving an intermediate result from all other nodes, which results in~${\gamma=1-\frac{1}{N} \approx 1}$. In this case, we have~$\beta\leq 0.404$, showing that the system can tolerate up to~$4040$ malicious nodes. Conversely, by setting~$\beta=0$, the lower bound of~$\gamma$ is~$0.192$, indicating that nodes can perform decoding after receiving~$1920$ intermediate results in an adversary-free system.

For a practical system to be robust in a network with~$\beta=0.3$, that is, 3000 nodes are malicious, we let~$\gamma=0.792$. That is, every honest node should perform decoding after receiving~$7920$ intermediate results to guarantee the correctness of verification results.

%malicious nodes can be tolerated, which accounts for~$\beta=40.23\%$ of the nodes.
\subsection{Low Latency in Limited Bandwidth}
As discussed in section Section~\ref{section:propagation}, our scheme requires
$${\frac{2(\sqrt{K}-1)}{D}+\log_{(D+1)}\frac{N}{K}}+1$$ rounds to deliver both the coded incoming and coded outgoing strips to every node, where the capacity~$D$ measures the amount of data that can be sent by one node in a constant time interval, measured in strips. This result shows that, the latency for a transaction to be verified grows sublinearly with the number of shards~$K$. It is worth noting that, this latency gain is achieved without high bandwidth requirement for nodes. In the first and second stage of the propagation, leader nodes cooperatively complete themselves, and each of them downloads amount of data equivalent to~$\sqrt{K}$ strips, which scales sublinearly with the number of shards~$K$. Later, non-leader nodes are completed in the third stage, and each of them is required to download amount of data equivalent to two strips, which is a constant independent of~$K$. 

\section{Future Research Directions}\label{section:future}
Our scheme supports cross-shard transactions and provides fast transaction confirmation (i.e., low latency), high throughput, and addresses the degree problem of the verification polynomials mentioned in Polyshard. Meanwhile, we suggest the following future research directions.
\begin{enumerate}
    \item Our scheme focuses on the cryptographic verification of transactions, and does not contain a countermeasure for double-spendings. Filling this gap is a significant direction of future researches. 
    \item We mainly focus on the security in the coded verification phase, and did not address the security problems in the strip propagation phase, as malicious nodes may undermine the verification result by sending erroneous strips. This problem can be seen as a special case of a decentralized encoding problem, which has recently been studied in~\cite{DistributedEncoding}.
    \item Our scheme adopts a simplified UTXO model, in which one transaction transfers one coin from one user to another. Removing this limitation and adopting a more practical UTXO model is also an interesting direction.
\end{enumerate}
%\clearpage

\clearpage

\end{document}